\documentclass[letterpaper, 10 pt, conference]{ieeeconf}  

\IEEEoverridecommandlockouts                              
\usepackage{mathtools}
\usepackage{amsmath}
\usepackage{amssymb}
\usepackage{mathrsfs}
\usepackage{graphicx}
\usepackage{subfigure}
\usepackage{color}
\usepackage{bm}
\usepackage{array}
\usepackage{algorithm}
\usepackage{algorithmic}

\makeatletter
\let\NAT@parse\undefined
\makeatother
\usepackage{hyperref}

\allowdisplaybreaks[4]

\newtheorem{asmp}{Assumption}

\newtheorem{thm}{Theorem}

\newtheorem{rmk}{Remark}
\newtheorem{defn}{Definition}

\newcommand{\RR}{\mathbb{R}}
\newcommand{\EE}{\mathbb{E}}
\newcommand{\PP}{\mathbb{P}}
\newcommand{\NN}{\mathbb{N}}

\newcommand{\mtcc}{\mathcal{C}}

\newcommand{\mtce}{\mathcal{E}}

\newcommand{\mtcg}{\mathcal{G}}

\newcommand{\mtcs}{\mathcal{S}}

\newcommand{\mtcv}{\mathcal{V}}
\newcommand{\mtcw}{\mathcal{W}}

\newcommand{\Let}{: =}

\newcommand{\bfl}{\mathbf{1}}
\newcommand{\bfo}{\mathbf{0}}
\newcommand{\bfx}{\mathbf{x}}
\newcommand{\tx}{\textup}
\newcommand{\tp}{\tx{T}}

\title{\LARGE \bf
Almost Exact Recovery in Gossip Opinion Dynamics\\ over Stochastic Block Models
}

\author{Yu Xing and Karl H. Johansson
\thanks{This work was supported by the Knut and Alice Wallenberg Foundation (Wallenberg Scholar Grant), the Swedish Research Council (Distinguished Professor Grant 2017-01078), the Swedish Foundation for Strategic Research (CLAS Grant RIT17-0046).}
\thanks{The authors are with Division of Decision and Control Systems, School of Electrical Engineering and Computer Science, KTH Royal Institute of Technology, and also with Digital Futures, Stockholm, Sweden. Email:
        {\tt\small \{yuxing2,kallej\}@kth.se}}
}

\begin{document}

\maketitle
\thispagestyle{empty}
\pagestyle{empty}

\begin{abstract}
We study community detection based on state observations from gossip opinion dynamics over stochastic block models (SBM). It is assumed that a network is generated from a two-community SBM where each agent has a community label and each edge exists with probability depending on its endpoints' labels. A gossip process then evolves over the sampled network. We propose two algorithms to detect the communities out of a single trajectory of the process. It is shown that, when the influence of stubborn agents is small and the link probability within communities is large, an algorithm based on clustering transient agent states can achieve almost exact recovery of the communities. That is, the algorithm can recover all but a vanishing part of community labels with high probability. In contrast, when the influence of stubborn agents is large, another algorithm based on clustering time average of agent states can achieve almost exact recovery. Numerical experiments are given for illustration of the two algorithms and the theoretical results of the paper.
\end{abstract}

\section{Introduction}
Networks exist ubiquitously in various fields such as computer science, biology, and sociology. It is common that nodes in a network connect densely within subgroups but sparsely in general. Such subgroups are referred to as communities~\cite{fortunato2016community}. 
Community detection is one of the central questions in network science and studies how to find communities of a network. Often only state dynamics evolving over the network are observable, rather than the network itself. Hence, a growing number of studies have been investigating community detection based on state observations~\cite{prokhorenkova2022less, peixoto2019network, wai2019blind, schaub2020blind, roddenberry2020exact, ramezani2018community}. 
Lacking network information makes community detection difficult, and it is still not clear how to detect communities based on a single trajectory of networked dynamics, which is considered in this paper.


\subsection{Related Work}
Community detection has been studied for two decades in multiple domains including physics and computer science~\cite{fortunato2016community,fortunato202220}.
Traditional methods apply agglomerative or divisive clustering to pairs of nodes with given weights~\cite{girvan2002community}. 
A popular concept for communities, called modularity, is introduced in~\cite{newman2004finding}. The modularity measures the quality of a given graph partition from a random partition. The Louvain method~\cite{blondel2008fast} is a renowned fast community detection algorithm based on modularity. A statistical approach modeling communities is to introduce generative network models that have planted community structures. A canonical model is the stochastic block model (SBM), in which each node has a pre-assigned community label and each edge exists with independent probability depending on its endpoints' labels. Spectral clustering and belief propagation methods are commonly used for the detection problem~\cite{abbe2017community}. Another detection approach is to execute dynamical processes over the network, for example, the Infomap algorithm~\cite{rosvall2008maps}. The authors in~\cite{morarescu2010opinion} introduces a bounded-confidence model, in which agents eventually form clusters coinciding with communities of the network.

In the aforementioned approaches, the network is assumed to be known. However, in practice it is likely the dynamic state data are available, instead of the network itself. As a result, there is a growing interest in community detection for networked systems based on state observations. Maximum likelihood methods applied to cascade data are introduced in~\cite{prokhorenkova2022less,ramezani2018community}. The paper~\cite{prokhorenkova2022less} also proposes a two-step procedure: first the underlying network is recovered and then agents are grouped from the network estimates. Blind community detection~\cite{wai2019blind,schaub2020blind,roddenberry2020exact} uses sample covariance matrices of agent states to recover the community structure. Estimating covariance matrices requires capturing a single snapshot from each of multiple trajectories. The paper~\cite{peixoto2019network} investigates learning the network topology and the community structure at the same time, for epidemics and an Ising model. 
The papers~\cite{xing2020community,xing2023community} consider a gossip model over a weighted graph with two communities, where agents within the same community have the same interaction probability, different from the interaction probability between communities. It is shown that the community structure can be recovered by clustering the state time average of the process. There is  a need to investigate how detection algorithms based on a single trajectory can be applied to general graphs.

In this paper we study the detection of communities from the gossip model with stubborn agents. The problem is related to recently increasing research of estimating network structure from social dynamics~\cite{ravazzi2021learning}. Network information is useful in applications, but directly collecting such data is hard because networks are topic specific~\cite{cowan2018could} and perturbed by noise~\cite{netrapalli2012learning}. Detecting communities for a coarse characterization of a network is a better choice than estimating all edges of that network, which is computationally expensive. The gossip update rule describes the stochastic nature of personal encounters and is a key building block of more complex opinion models \cite{proskurnikov2017tutorial}. In social network modeling, media, influential bloggers, and opinion leaders can be seen as stubborn agents, whose existence may have a great effect on the dynamics and result in opinion fluctuation~\cite{acemouglu2013opinion}.

\subsection{Contributions}\label{sec_sub_contribution}
This paper studies community detection based on a single trajectory of gossip opinion dynamics over a two-community SBM. Two detection algorithms are proposed. The first algorithm (Algorithm~\ref{alg_1}) is based on applying the $k$-means algorithm to agent states in a given transient time interval. It shown in Theorem~\ref{alg_1} that, if the influence of stubborn agents is small and the link probability within communities is large, the algorithm can recover all but a vanishing proportion of community labels of the SBM with high probability (i.e., almost exact recovery). The time interval depends on the relative magnitude of link probability within and between communities. The second algorithm (Algorithm~\ref{alg_2}) deals with the case where the influence of stubborn agents is large. The algorithm computes the time average of agent states and returns community estimates by clustering the time average after a given time step. Theorem~\ref{thm_2} states that this algorithm can also achieve almost exact recovery.

The results generalize the earlier works~\cite{xing2020community,xing2023community}, in which the graph with two communities is deterministic and fixed. The difficulty lies in theoretically characterizing the relation between agent states and the community structure of the SBM. We verify almost exact recovery by using new concentration results developed in~\cite{xing2022transient,xing2023concentration}.

The results show that a single trajectory of the gossip model can be enough for achieving almost exact recovery. Also, the proposed algorithm based on transient states indicates that excitation from stubborn agents may not be necessary to guarantee recovery. The analysis framework provides insight into design and analysis of community detection algorithm based on state observations. Given a process evolving over a structured network, we can first study how the community structure influences the dynamics, and then exploit the obtained properties to design detection algorithms.

\subsection{Outline}
Section~\ref{sec:pre} introduces the SBM and the gossip model. Section~\ref{sec:problm} defines almost exact recovery of the SBM and formulates the problem. Section~\ref{sec:mainr} provides two algorithms and their performance analysis. Numerical experiments are presented in Section~\ref{sec:num}. Section~\ref{sec:concl} concludes the paper. 

\noindent\textbf{Notation.} Let $\mathbb{R}^n$, $\mathbb{R}^{n\times m}$, and $\mathbb{N}$ be the $n$-dimensional Euclidean space, the set of $n\times m$ real matrices, and the set of nonnegative integers, respectively. Denote $\NN_+ = \NN\setminus\{0\}$. For $x\in \RR$, $\log x$ is the natural logarithm of $x$. Denote the $n$-dimensional all-one vector and the  unit vector with $i$-th entry being one by $\mathbf{1}_n$ and $e_i^{(n)}$, respectively. The superscript $(n)$ is omitted if the context is clear. $I_n$ is the $n\times n$ identity matrix, and $\bfl_{m,n}$ ($\bfo_{m,n}$) is the $m\times n$ all-one (all-zero) matrix. Denote the Euclidean norm of a vector and the spectral norm of a matrix by $\|\cdot\|$.
For $x\in \mathbb{R}^n$, $x_i$ is its $i$-th entry, and for $A \in \mathbb{R}^{n\times n}$, $a_{ij}$ or $[A]_{ij}$ is its $(i,j)$-th entry.
The cardinality of a set $\mtcs$ is $|\mtcs|$. The function $\mathbb{I}_{[\textup{property}]}$ is the indicator function, which is one if the property in the bracket holds, and is zero otherwise. Denote the probability of an event $A$ by $\PP\{A\}$ and the expectation of a random vector $X$ by $\mathbb{E}\{X\}$.
For real numbers  $a(n)$ and $b(n) > 0$, $n\in\NN$, denote $a(n) = O(b(n))$ if $|a(n)| \le C b(n)$ for all $n\in \NN$ and some $C > 0$, $a(n) = o(b(n))$ if $|a(n)|/b(n) \to 0$. If further $a(n) > 0$, $n\in\NN$, denote $a(n) = \omega(b(n))$ if $b(n) = o(a(n))$, $a(n) = \Omega(b(n))$ if $b(n) = O(a(n))$, and $a(n) = \Theta(b(n))$ if both $a(n) = O(b(n))$ and $a(n) = \Omega(b(n))$. 
We will use subscripts to emphasize the dependence on $n$, for example, $a(n) = o_n(b(n))$. Denote $x\vee y \Let\max\{x,y\}$ and $x\wedge y\Let \min\{x,y\}$, $x,y \in \RR$. An undirected graph $\mtcg = (\mtcv, \mtce, A)$ has the agent set $\mtcv$, the edge set $\mtce$, and the adjacency matrix $A = [a_{ij}]$ with $a_{ij} = 1$ ($a_{ij} = 0$) if $\{i,j\} \in \mtce$ ($\{i,j\} \not\in \mtce$). The degree of~$i \in \mtcv$ is $d_i = \sum_{j \in \mtcv} a_{ij}$.

\begin{figure*}[tbp]
  \centering
  \includegraphics[scale=0.32]{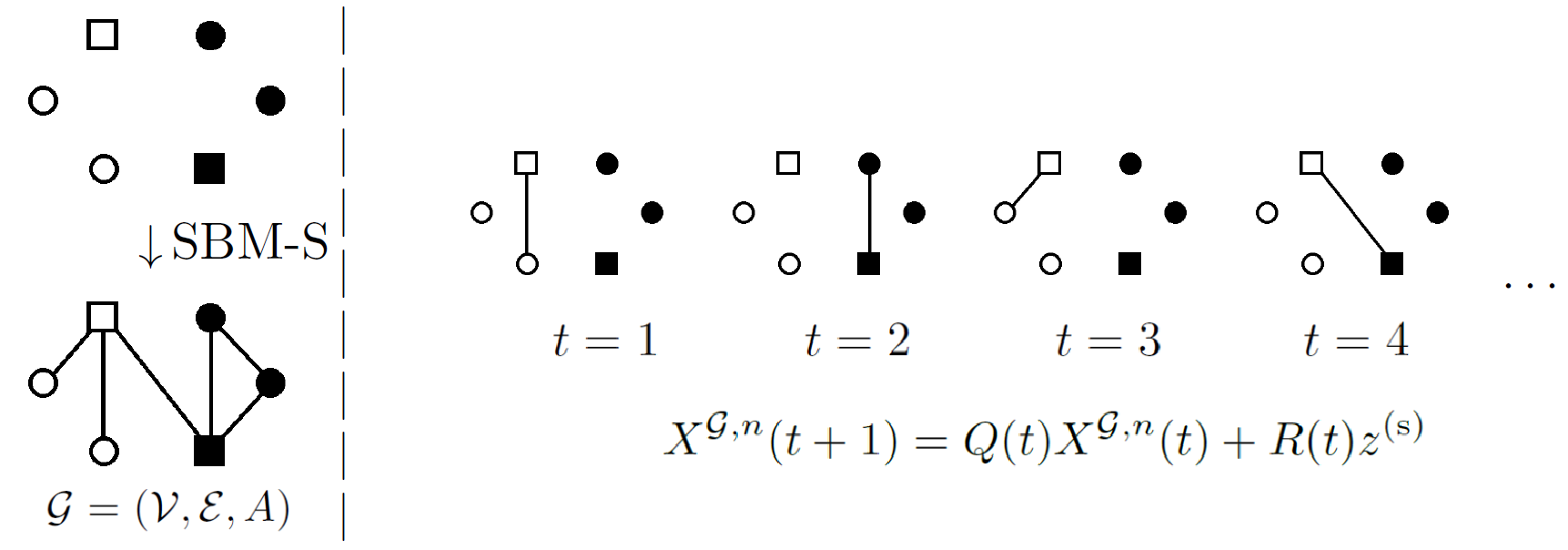}
  \caption{\label{fig:illustration}Illustration of the gossip model over an SBM-S. On the left side of the figure, a network is generated from the SBM-S and then fixed. Circles and squares represent regular and stubborn agents, respectively. The black and white are two communities.
  On the right, the gossip model evolves over the generated network. An edge is selected at a time.}
  
\end{figure*}

\section{Preliminaries}\label{sec:pre}
\subsection{Two-Community SBM}\label{subsec:sbm}

In this subsection, we define a two-community SBM. The SBM characterizes the community structure of real networks. For a graph $\mtcg$, we assume that its agent set $\mtcv$ can be represented by the union of two disjoint sets $\mtcv_{\tx{r}1}$ and $\mtcv_{\tx{r}2}$. Let the vector $\mtcc \in \{1,2\}^n$ be such that $\mtcc_i = 1$ if $i \in \mtcv_{\tx{r}1}$ and $\mtcc_i = 2$ if $i \in \mtcv_{\tx{r}2}$. In other words, agents in $\mtcv_{\tx{r}1}$ (in $\mtcv_{\tx{r}2}$) have the label~$1$ (label~$2$). We call $\mtcv_{\tx{r}1}$ and $\mtcv_{\tx{r}2}$ the communities of the graph and $\mtcc$ the community structure.
\begin{defn}[SBM]\label{def:sbm}
	Let $n\in \NN_+$ be an even number, $\bm{l} = [l_\tx{s}~l_\tx{d}]^\tp = [l_\tx{s}(n)~l_\tx{d}(n)]^\tp \in (0,1)^2$. The $\tx{SBM}(n,\bm{l})$ is a random graph. The SBM assigns agents $1$, $\dots$, $n/2$ (agents $1+n/2$, $\dots$, $n$) with label~$1$ (label~$2$). Then it generates an undirected graph $\mtcg = (\mtcv,\mtce,A)$, without self-loops, by independently adding $\{i,j\}$ with $i\not=j$ to $\mtce$ with probability $l_{\tx{s}} \mathbb{I}_{[\mtcc_i = \mtcc_j]} + l_{\tx{d}} \mathbb{I}_{[\mtcc_i \not= \mtcc_j]}$.\hfill\QED
\end{defn}

From this definition, we know that a graph generated from an SBM has communities $\mtcv_{\tx{r}1} = \{1, \dots, n/2\}$, $\mtcv_{\tx{r}2} = \{1+n/2, \dots, n\}$, and community structure $\mtcc = [\bfl_{n/2}^\tp~ 2\bfl_{n/2}^\tp]^\tp$.

\begin{rmk}
    The size of communities can be a random variable instead of a fixed number as assumed in the definition (see Remark~3 of~\cite{abbe2017community}).\hfill\QED
\end{rmk}

\subsection{Gossip Model with Stubborn Agents}\label{subsec:opinionmodel}

This subsection introduces the gossip model with stubborn agents. The model captures random personal encounters. The gossip model is a random process over an undirected graph $\mtcg = (\mtcv, \mtce, A)$. The agent set $\mtcv = \mtcv_{\tx{r}} \cup \mtcv_{\tx{s}}~ \tx{(disjoint)}$  has both regular and stubborn agents. For convenience, denote $\mtcv_{\tx{r}} = \{1, \dots, n_{\tx{r}}\}$ and $\mtcv_{\tx{s}} = \{1+n_{\tx{r}}, \dots, n_{\tx{s}} + n_{\tx{r}}\}$, so $|\mtcv| = n = n_{\tx{r}}+n_{\tx{s}}$. Here $n_{\tx{r}}$ ($n_{\tx{s}}$) is the number of regular (stubborn) agents, and $n$ is the network size.
Regular agents have opinion vector $X(t) \in \RR^{n_{\tx{r}}}$ at time $t \in \NN$, and $X_i(t)$ is the opinion of the agent~$i$. Stubborn agents have opinion vector $z^{(\tx{s})} \in \RR^{n_{\tx{s}}}$ with $z_j^{(\tx{s})}$ being the opinion of the stubborn agent~$j+n_{\tx{r}}$. An edge $\{i,j\}$ is chosen at each time $t$ independent of previous updates, with an interaction probability $w_{ij} = w_{ji} = a_{ij}/\alpha$, where $\alpha = \sum_{i=1}^n \sum_{j=i+1}^n a_{ij} = |\mtce|$ and $W = [w_{ij}] \in \RR^{n\times n}$ is the interaction probability matrix. The two corresponding agents then update according to the following rule: If~$i$ and~$j$ are regular, then $X_i(t+1) = X_j(t+1) = (X_i(t) + X_j(t))/2.$
If one of them is stubborn, for example $j$, then $j$ does not update, but $X_i(t+1) = (X_i(t) + z^{(\tx{s})}_j)/2$. All other agents keep their states at $t$. The compact form of the update is
\begin{align}\label{eq:gossipmodelo}
	X(t+1) = Q(t) X(t) + R(t) z^{(\tx{s})}, 
\end{align}
with $\{[Q(t)~R(t)]\}$  a sequence of i.i.d. random matrices. With probability~$w_{ij}$ if $i, j \in \mtcv_{\tx{r}}$ then $[Q(t),~R(t)] = [I_{n_{\tx{r}} } - \frac12 (e_i^{(n_{\tx{r}})} - e_j^{(n_{\tx{r}})} )(e_i^{(n_{\tx{r}})}  - e_j^{(n_{\tx{r}})}) ^\tp,~\bfo_{n_{\tx{r}},n_{\tx{s}}}]$, and if $i \in \mtcv_{\tx{r}},  j \in \mtcv_{\tx{s}}$ then $[I_{n_{\tx{r}} } - \frac12 e_i^{(n_{\tx{r}})}(e_i^{(n_{\tx{r}})})^\tp,~\frac12 e_i^{(n_{\tx{r}})} (e_j^{(n_{\tx{s}})})^\tp]$.
Here we replace $e^{(n_{\tx{s}})}_{j-n_{\tx{r}}}$ with $e_j^{(n_{\tx{s}})}$ for $j \in \mtcv_{\tx{s}}$ for convenience.

\subsection{SBM with Stubborn Agents}\label{subsec:randomwithstubborn}
To define a gossip model over an SBM, in this section we introduce an SBM with stubborn agents based on Definition~\ref{def:sbm}.
\begin{defn}[SBM with stubborn agents, SBM-S]\label{def:sbm_s}
	Let the number of regular agents $n_{\tx{r}} \in \NN_+$ be an even number, the number of stubborn agents $n_{\tx{s}} \in \NN_+$, the network size $n = n_{\tx{r}} + n_{\tx{s}}$, the link probability between regular agents $\bm{l} = [l_\tx{s}~l_\tx{d}]^\tp = [l_\tx{s}(n)~l_\tx{d}(n)]^\tp \in (0,1)^2$, and the link probability matrix between regular and stubborn agents $L^{(\tx{s})} = [l_{ij}^{\tx{(s)}}] = [l_{ij}^{\tx{(s)}}(n)] \in [0,1)^{n_{\tx{r}}\times n_{\tx{s}}}$. The $\tx{SBM-S}(n_{\tx{r}},n_{\tx{s}},\bm{l},L^{\tx{(s)}})$ is a random graph which first generates an undirected graph on the $n_{\tx{r}}$ regular agents from $\tx{SBM}(n_{\tx{r}},\bm{l})$ and then add each edge $\{i,j\}$ to the graph with probability $l_{ij-n_{\tx{r}}}^{\tx{(s)}}$ for $i\in \mtcv_{\tx{r}} = \{1,\dots,n_{\tx{r}}\}$ and $j\in \mtcv_{\tx{s}} = \{1+n_{\tx{r}}, \dots, n\}$.\hfill\QED
\end{defn}

The SBM-S includes stubborn agents in a network. The probability $l_{ij}^{(\tx{s})}$ measures the possibility of the regular agent~$i$ connected to the stubborn agent~$j+n_{\tx{r}}$. Let $r_0\Let n_{\tx{r}}/n \in (0,1)$ be the proportion of the regular, and $s_0 \Let n_{\tx{s}}/n \in (0,1)$ that of the stubborn. Hereafter by a gossip model we mean a gossip model that evolves over a sample graph generated by an SBM-S (see Fig.~\ref{fig:illustration} for an illustration).

\section{Problem Formulation}\label{sec:problm}
We study how to detect the community structure of an SBM-S out of state observations. To measure the performance of a detection algorithm, denote the accuracy of an estimate $\hat{\mtcc}$ of the community structure $\mtcc$ by 
\begin{align}\label{eq_accdefn}
    \tx{Acc}(\mtcc,\hat{\mtcc}) \Let \frac1n \max\bigg\{\sum_{i=1}^n \mathbb{I}_{[\mtcc_i = \hat{\mtcc}_i]}, \sum_{i=1}^n \mathbb{I}_{[\mtcc_i = 3-\hat{\mtcc}_i]} \bigg\}.
\end{align}
{\color{black}The first summation in~\eqref{eq_accdefn} represents the number of identical entires between $\mtcc$ and $\hat{\mtcc}$. Note that $\mtcc_i \in \{1,2\}$, so $3-\hat{\mtcc}_i$ swaps the community label of~$i$, and the second term in~\eqref{eq_accdefn} represents the number of identical entires between $\mtcc$ and the swapped estimated labels. That is, the accuracy is defined up to a permutation of labels.}
Now define almost exact recovery of an algorithm detecting communities in SBMs as follows~\cite{abbe2017community}.
\begin{defn}
    For an SBM with $n$ agents and a community structure $\mtcc$, suppose that a detection algorithm outputs an estimation of the community structure $\hat{\mtcc}$. We say that the algorithm achieves almost exact recovery, if 
    \begin{align*}
        \qquad\qquad\PP \{ \tx{Acc}(\mtcc,\hat{\mtcc}) = 1 - o_n(1) \} = 1 - o_n(1).\tx{\qquad\quad~\QED}
    \end{align*}
\end{defn}
\begin{rmk}
    Almost exact recovery means that the algorithm can detect most community labels (up to a permutation) except for a vanishing part with probability approaching one, as the network size increases. \hfill\QED
\end{rmk}

The problem considered in this paper is as follows:

\emph{Problem:} For an SBM-S and the gossip model over this SBM-S, given a trajectory of the model, $\{X(0),\dots,X(t),\dots\}$, propose algorithms that use the trajectory data to achieve almost exact recovery of the regular agents' community structure.\hfill\QED

In the next section, we will show that almost exact recovery can be achieved from clustering transient agent states (Theorem~\ref{thm_1}). If the influence of stubborn agents is large enough, then almost exact recovery can be achieved from clustering state time averages of the model (Theorem~\ref{thm_2}).

\section{Detection Algorithms and Performance}\label{sec:mainr}
In this section we propose two detection algorithms using a single trajectory of the gossip model, and show that the algorithm using transient states (Algorithm~\ref{alg_1}) achieves almost exact recovery (Theorem~\ref{thm_1}), and so does the algorithm using state time average (Algorithm~\ref{alg_2} and Theorem~\ref{thm_2}).

First, we introduce the following assumptions.

\begin{asmp}\label{asmp_1}
    Suppose that the following hold.\\
    (i) There exists $l^{(\tx{s})} \ge 0$ such that $\sum_{1\le j \le n_{\tx{\tx{s}}}} l_{ij}^{(\tx{s})} = l^{(\tx{s})} = l^{(\tx{s})}(n)$ for all $i \in \mtcv_{\tx{r}}$.\\
    (ii) The  vectors $X(0)$ and $z^{(\tx{s})}$ are deterministic, and satisfy that $|X_i(0)| \le c_x$ and $|z^{(\tx{s})}_j| \le c_x$, for all $1\le i \le n_{\tx{r}}$ and $1\le j\le n_{\tx{s}}$, and some constant $c_x>0$.\\
    (iii) The proportion of regular agents $r_0$ is a constant and $r_0 n$ is an even number. \hfill\QED
\end{asmp}
\begin{rmk}
    In~(i), $l^{(\tx{s})}$ represents the total influence of stubborn agents on one regular agent. We impose the first assumption for simplicity. {\color{black}It is possible to analyze the general problem by using matrix perturbation theory and the upper and lower bounds of $\sum_{1\le j \le n_{\tx{s}}} l_{ij}^{(\tx{s})}$.} The assumption~(ii) implies that the process is bounded. Finally, {\color{black} note that $r_0$ in (iii) is fixed but can be any number in $(0,1)$. The case where $r_0n$ is an odd number can be studied similarly to the case of even numbers.}\hfill\QED
\end{rmk}

We also need the following assumption for the link probabilities of the SBM-S.
\begin{asmp}\label{asmp:degree}It holds that $l_{\tx{s}} = \omega((\log n)/n)$ and $l^{(\tx{s})} = \omega(\log n)$.\hfill\QED
\end{asmp}
\begin{rmk}
    The assumption show that we consider the logarithm regima of SBMs, where an SBM generates connected graphs~\cite{abbe2017community}. {\color{black}
    The condition $l^{(\tx{s})} = \omega(\log n)$ is required because the influence of stubborn agents needs to be large enough to ensure concentration. This condition may be removed, which is left to future work.} \hfill\QED
\end{rmk}

\begin{algorithm}[t]
\caption{(Recovery Based on Transient States)}
\label{alg_1}
\small \textbf{Input:} Trajectory $\{X(t), t\in \NN\} $.\\
\textbf{Output:}~Community estimate~$\hat{\mathcal{C}}$.
\begin{algorithmic}[1]
\STATE{Select a vector $X(t)$ with $t \in (\Theta(n\log n), o(nl_{\tx{s}}/l_{\tx{d}}))$.}
\STATE{Obtain $\hat{\mathcal{C}}$ by applying $k$-means to $X(t)$.}
\end{algorithmic}
\end{algorithm}

We now propose the first detection algorithm (Algorithm~\ref{alg_1}), based on the intuition that regular agents within communities may have similar states when the process has not evolved too long, if the influence of stubborn agents is small. Although simple, proving the recovery result needs analysis of concentration of $X(t)$, which is not trivial (see~\cite{xing2022transient} for the details). In application, $l_{\tx{s}}$ and $l_{\tx{d}}$ are unknown, so we can use $X(t)$ with $t = \Theta(n\log n)$ for clustering. Improvement of the algorithm will be studied in the future. The recovery performance of Algorithm~\ref{alg_1} is given in the following theorem.
\begin{thm}\label{thm_1}
    Suppose that Assumptions~\ref{asmp_1} and~\ref{asmp:degree} hold. If $l_{\tx{s}}/(\log n) = \omega(l_{\tx{d}})$, $(l_{\tx{s}}\wedge l_{\tx{d}})n/(\log n) = \omega( l^{(\tx{s})})$, $l_{\tx{d}} = \omega([l_{\tx{s}} (\log n)/n]^{1/2})$ and $\sum_{i \in \mtcv_{\tx{r}1}} X_i(0) \not = \sum_{j\in\mtcv_{\tx{r}2}} X_j(0)$, then Algorithm~\ref{alg_1} achieves almost exact recovery.\hfill\QED
\end{thm}
\quad\quad\emph{Proof Sketch:}
 The main idea of the proof is to compare the gossip dynamics over an SBM-S with a gossip dynamics over an averaged graph. The averaged graph $\bar{\mtcg} = (\mtcv, \bar{\mtce}, \EE\{A\})$ is obtained by averaging all possible graphs $\mtcg = (\mtcv, \mtce, A)$ generated from the SBM-S, where $\EE\{A\}$ is the weighted adjacency matrix and $[\EE\{A\}]_{ij} = l_{\tx{s}}\mathbb{I}_{[\mtcc_i = \mtcc_j]} + l_{\tx{d}}\mathbb{I}_{[\mtcc_i \not= \mtcc_j]}$ for $i\not=j\in \mtcv_{\tx{r}}$. A gossip dynamics taking place over this averaged graph has an interaction probability matrix $\mtcw = \EE\{A\}/\EE\{\alpha\}$. 

As shown in Theorem~1 of~\cite{xing2022transient}, agent states of the gossip dynamics over the averaged graph are close to opinion averages within their corresponding communities, under the assumptions of the current theorem. By using Bernstein concentration inequalities~\cite{gross2011recovering,vershynin2018high}, we can show that the variance of agent states $X(t)$ of the original gossip dynamics is close to that over the averaged graph. Hence agent states of the original gossip dynamics are also close to the corresponding community averages. 
Note that the $k$-means problem on the real line can be solved optimally~\cite{blum2020foundations}. By a counting argument, it is possible to show that the optimal partition of $X(t)$ is the same as the community structure, except for a vanishing set of agents. The conclusion then follows.
\hfill\QED

\begin{algorithm}[t]
\caption{(Recovery Based on State Time Average)}
\label{alg_2}
\small \textbf{Input:} Trajectory $\{X(t), t\in \NN\}$,  time step for clustering $T \in \NN_+$.\\
\textbf{Output:} Community estimate $\hat{\mathcal{C}}$.
\begin{algorithmic}[1]
\STATE{Set $S(0) = X(0)$. }
\FOR{$t = 1, \dots, T$}
\STATE{Compute
\begin{align*}
    S(t) &= \frac{t}{t+1} S(t-1) + \frac{1}{t+1} X(t).
\end{align*}
} 
\ENDFOR
\STATE{Obtain $\hat{\mtcc}$ by applying $k$-means to $S(T)$.}
\end{algorithmic}
\end{algorithm}

\begin{rmk}
    The theorem indicates that, if the influence of stubborn agents is small and the link probability within communities is large, then it is possible to recover most labels based on agent states at a time step smaller than $nl_{\tx{s}}/l_{\tx{d}}$. The possible time interval depends on the relative magnitude of link probability within and between communities. More careful analysis can remove the term $\log n$ in the condition. {\color{black}The condition $(l_{\tx{s}}\wedge l_{\tx{d}})n/(\log n) = \omega( l^{(\tx{s})})$ implies that link probability between regular agents grows faster with $n$ than the influence of stubborn agents. This represents the maximum allowable effect of stubborn agents on regular agents for guaranteeing almost exact recovery.}\hfill\QED
\end{rmk}

When the influence of stubborn agents is large, the process can reach its steady states quickly~\cite{xing2022transient}, so it is difficult to find an interval suggested by Algorithm~\ref{alg_1}. For this case, we compute the state time average and cluster this vector to obtain an estimate of the community structure (Algorithm~\ref{alg_2}). In the algorithm, the state time average $S(t) = (\sum_{j=0}^{t-1} X(i))/t$ is computed recursively. In practice, the link probabilities are unknown, so it may be hard to decide how to set $T$ in Algorithm~\ref{alg_2}. A possible way is to cluster $S(t)$ for every $t \in \NN$ and to terminate the process when the change of community estimates is below a given threshold.

To guarantee exact recovery of Algorithm~\ref{alg_2} and simplify analysis, we need the following technical assumption. 
\begin{asmp}\label{asmp_z}
    For the SBM-S and the stubborn agent states $z^{\tx{(s)}}$, it holds that $L^{\tx{(s)}} z^{\tx{(s)}} = [\zeta_1 \bfl_{n_{\tx{r}}/2}^\tp~\zeta_2 \bfl_{n_{\tx{r}}/2}^\tp]^\tp$ with $\zeta_1,\zeta_2 \in \RR$ and $\zeta_1 \not=\zeta_2$.\hfill\QED
\end{asmp}
\begin{rmk}
    The assumption means that the weighted average of stubborn-agent states for each regular agent is the same within communities, and the averages are different between the two communities. We introduce this condition to ensure that the expected steady agent states between communities are different. Thus the community structure can be recovered based on these different values. The condition can be replaced by a bound controlling the difference between $[L^{\tx{(s)}} z^{\tx{(s)}}]_{i}$ and $[L^{\tx{(s)}} z^{\tx{(s)}}]_{j}$ for $i\in \mtcv_{\tx{r}1}$ and $j \in \mtcv_{\tx{r}2}$.\hfill\QED
\end{rmk}

The almost exact recovery by Algorithm~\ref{alg_2} is stated in the following theorem.
\begin{thm}\label{thm_2}
    Suppose that Assumptions~\ref{asmp_1}--\ref{asmp_z} hold. If $l_{\tx{s}} = \omega(l_{\tx{d}})$ and $l^{\tx{(s)}} \ge l_{\tx{s}}n$, then Algorithm~\ref{alg_2} achieves almost exact recovery if $T = \Omega(n^3l^{\tx{(s)}})$.\hfill\QED
\end{thm}
\quad\quad\emph{Proof Sketch:}
Similar to the proof of Theorem~\ref{thm_1}, here we compare the expected final opinions $\bfx^{\mtcg,n} \Let  (I - \EE\{Q(t)\})^{-1} \EE\{R(t)\} z^{(\tx{s})}$ with the expected final opinions of the gossip dynamics over the averaged graph. Theorem~4.3 of~\cite{xing2023concentration} guarantees that with high probability the difference between these two vectors can be bounded by a vanishing error. Moreover, Theorem~4.11 and Remark~4.12 of~\cite{xing2023concentration} yield that the state time average $S(t)$ is close to $\bfx^{\mtcg,n}$  for $t = \Omega(n^3l^{\tx{(s)}})$. Hence applying $k$-means to $S(t)$ yields the desired partition with high probability.
\hfill\QED
\begin{rmk}
    The theorem indicates that almost exact recovery can be achieved if the stubborn agent influence and the link probability between communities are large enough. The bound for the clustering time $\Theta(n^3l^{\tx{(s)}})$ can be relaxed, as observed in the numerical experiments  in Section~\ref{sec:num}. \hfill\QED
\end{rmk}

\begin{figure}[t]
	\centering
	\includegraphics[scale=0.35]{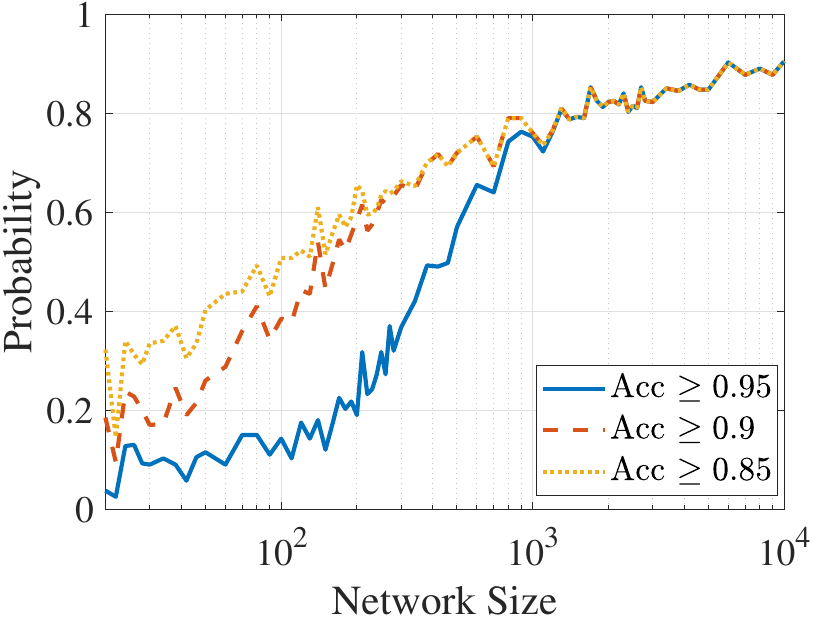}
	\caption{\label{fig_almost} Accuracy of Algorithm~\ref{alg_1}.}	
\end{figure}

\section{Numerical Simulation}\label{sec:num}
In this section we present numerical experiments to validate the main results. We test the proposed algorithms for gossip dynamics over an SBM-S and dynamics over a real network.

To generate an SBM-S with $n$ agents, set the proportion of regular agents to be $r_0 = 0.9$, so the proportion of stubborn agents is $s_0 = 0.1$. Recall the link probability within communities $l_{\tx{s}}$ and the link probability between communities $l_{\tx{d}}$. For the link probability between regular and stubborn agents, let
\begin{align*}
    L^{\tx{(s)}} = \begin{bmatrix}
        l_1^{\tx{(s)}} \bfl_{n_{\tx{r}}/2,n_{\tx{s}}/2} & \bfo_{n_{\tx{r}}/2,n_{\tx{s}}/2} \\
        \bfo_{n_{\tx{r}}/2,n_{\tx{s}}/2} & l_1^{\tx{(s)}} \bfl_{n_{\tx{r}}/2,n_{\tx{s}}/2}
    \end{bmatrix}.
\end{align*}
Intuitively this matrix shows that the stubborn agents also form two communities, and each stubborn community influences only one regular community, with link probability $l_1^{\tx{(s)}}$. In this way we define an $\tx{SBM-S}(n_{\tx{r}},n_{\tx{s}},\bm{l},L^{\tx{(s)}})$, where $\bm{l} = [l_{\tx{s}}~l_{\tx{d}}]^\tp$. 
For the gossip model, we set the states of the first half of stubborn agents to be $1$ and the other half to be $-1$. 

\begin{figure}[t]
	\centering
	\includegraphics[scale=0.35]{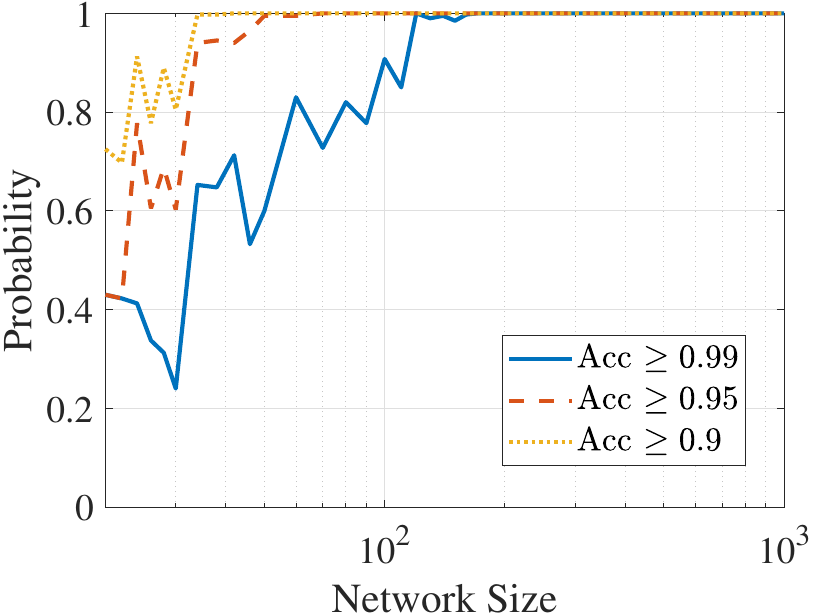}
	\caption{\label{fig_exact} Accuracy of Algorithm~\ref{alg_2}.}	
\end{figure}

\begin{figure*}[t]
    \centering
    \subfigure[\label{zachary_network}The community structure of Zachary's karate club network. Red squares and green triangles show two communities.]{\qquad\qquad
    \includegraphics[width=0.28\linewidth]{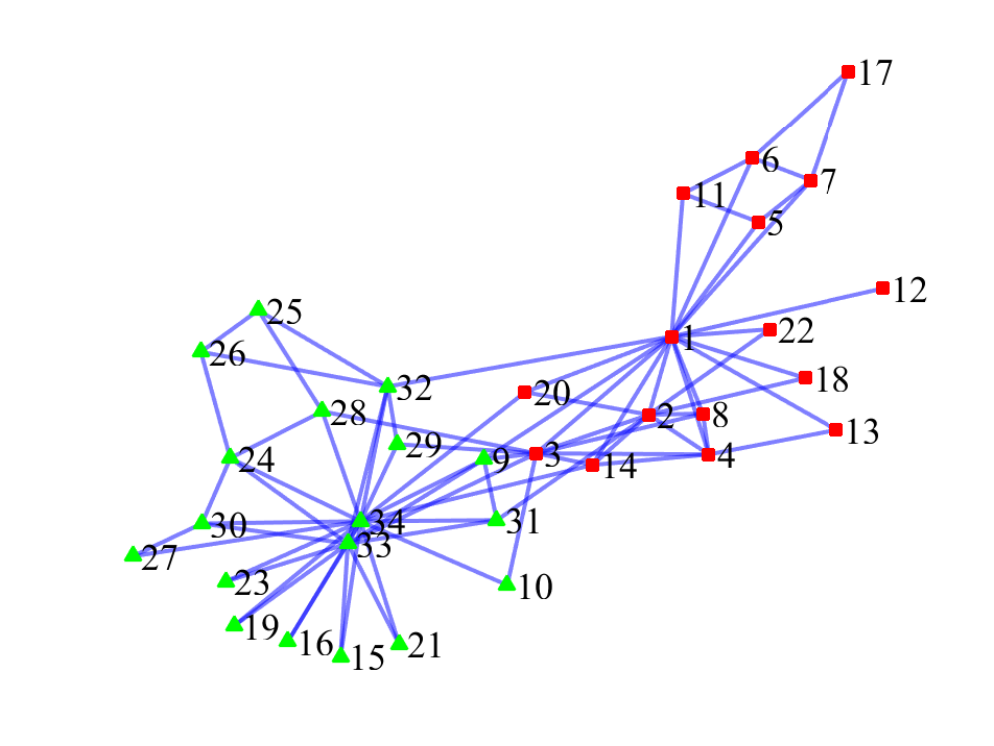}\qquad\qquad
    }~
    \subfigure[\label{zachary_accuracy}Accuracy of Algorithms~\ref{alg_1} and~\ref{alg_2} for community detection in  gossip dynamics over Zachary's karate club network.]{\qquad\qquad
    \includegraphics[width=0.28\linewidth]{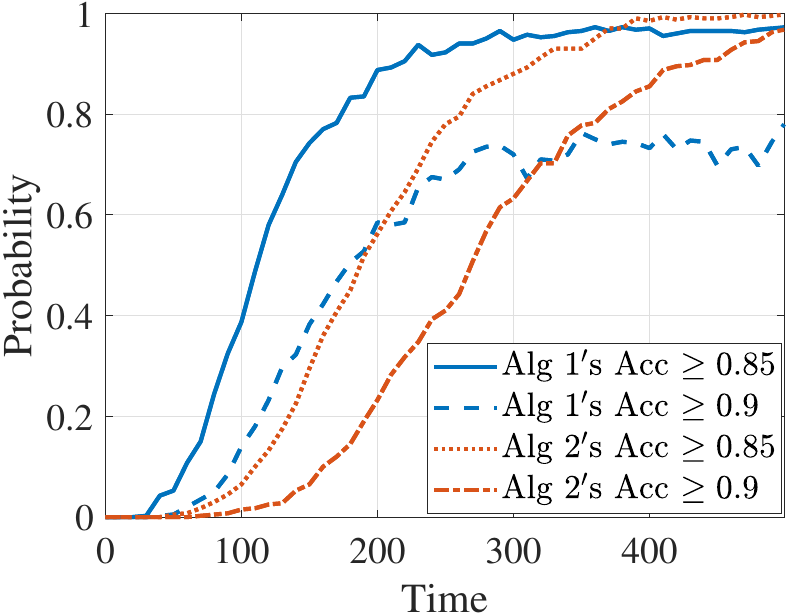}\qquad\qquad
    }
    \caption{\label{fig_2}Numerical experiment over Zachary's karate club network.}
\end{figure*}

To show the almost exact recovery of Algorithm~\ref{alg_1}, we set $l_{\tx{s}} = (\log n)^{2.5}/n$, $l_{\tx{d}} = (\log n)/n$, and $l_1^{\tx{(s)}} = \log n/n$. Generate the initial agent states in $\mtcv_{\tx{r}1}$ independently from the uniform distribution on $(-1,0)$, and those in $\mtcv_{\tx{r}2}$ independently and uniformly from $(0,1)$. Then run the gossip dynamics over the SBM-S for different $n$ from $10$ to $10^4$. {\color{black}For each $n$, $20$ graph samples of the SBM-S are generated and for each graph sample $20$ trajectories are collected.} Algorithm~\ref{alg_1} is applied to $X(t)$ with $t = [n\log n]$ for each trajectory, where $[\cdot]$ is the rounding function. {\color{black}Recall the accuracy of an algorithm is the proportion of correctly detected community labels up to a permutation, as given in~\eqref{eq_accdefn}.} Fig.~\ref{fig_almost} shows that the probability of the algorithm achieving a given accuracy increases with the number of agents. In addition, the accuracy of the algorithm also increases with the network size, when considering a given probability. These observations indicate that the algorithm achieves almost exact recovery.

For the almost exact recovery of Algorithm~\ref{alg_2}, we set set $l_{\tx{s}} = (\log n)^{2}/n$, $l_{\tx{d}} = (\log n)/n$, and $l_1^{\tx{(s)}} = (\log n)^{2.5}/n$. Generate the initial states of all regular agents independently from the uniform distribution on $(-1,1)$. Run the gossip model over the SBM-S for different $n$ from $10$ to $10^3$. As previously, {\color{black}$20$ graph samples are generated for each $n$ and $20$ trajectories are collected for each graph sample.} Algorithm~\ref{alg_2} is applied to each trajectory with the final step $T = [n(\log n)^{2.5}]$. Almost exact recovery of the algorithm is shown in Fig.~\ref{fig_exact}. The result also indicates that Algorithm~\ref{alg_2} performs better than Algorithm~\ref{alg_1}, and Algorithm~\ref{alg_2} can recover all labels even for relatively small $n$. This may be because stubborn agents produce more excitation. Also note that $T$ is much smaller than the condition given in Theorem~\ref{thm_2}, suggesting the possiblity of improving the results.

{\color{black}To test the performance of the proposed algorithms, we also conduct a numerical experiment over Zachary’s karate club network, shown in Fig.~\ref{zachary_network}. This network with two communities has been widely used as a benchmark in community detection. In the network, an edge represents interactions between agents. We assume that gossip dynamics take place over the network and we can observe only agent opinions but not interactions or the network. Agents $1$ and $34$ are leaders of the two communities, so they are set to be stubborn agents holding opinions $1$ and $-1$, respectively. Other agents are regular ones with initial opinions uniformed generated from $(-1,1)$. During evolution, an edge in Fig.~\ref{zachary_network} is uniformly randomly selected at each time. We run the model for $400$ times and apply Algorithms~\ref{alg_1} and~\ref{alg_2} to every time step of each trajectory. Fig.~\ref{zachary_accuracy} shows that Algorithm~\ref{alg_1} achieves high accuracy with high probability at the initial phase of the process, but Algorithm~\ref{alg_2} performs better when the process reaches steady state. The experiments indicate the applicability of both algorithms.}

\section{Conclusion}\label{sec:concl}
We studied community detection based on state observations from gossip opinion dynamics over a two-community SBM. Two algorithms were proposed and both of them use only a single trajectory of the process. When the influence of stubborn agents is small and the link probability within communities is large, it was shown that the algorithm based on clustering transient agent states can achieve almost exact recovery. In contrast, when the influence of stubborn agents is large, the algorithm based on clustering state time average can achieve almost exact recovery. Future work includes investigating the general SBM case and generalizing the algorithm to other networked dynamics.

\bibliographystyle{ieeetr}
\bibliography{biblio.bib}

\end{document}